\documentstyle[twocolumn,pra,aps]{revtex}

\begin{document}

\bibliographystyle{prsty}

\draft


\tighten

\title{Thermal photon statistics in laser light above threshold}

\author{Holger F. Hofmann and Ortwin Hess}
\address{Institut f\"ur Technische Physik, DLR\\
Pfaffenwaldring 38--40, D--70569 Stuttgart, Germany}

\date{\today}

\maketitle

\begin{abstract}
We show that the reduction in photon number fluctuations at 
laser threshold often cited as a fundamental laser property does not
occur in small semiconductor lasers.  The conventional theory of threshold
noise is not valid in lasers with a spontaneous emission 
factor larger than $10^{-8}$. If the spontaneous emission factor is larger
than $10^{-4}$, the photon number 
statistics even remain thermal far above threshold. We therefore conclude
that the reduction in photon number fluctuations is not a fundamental
laser property but rather a matter of size and the corresponding 
relative importance of quantum fluctuations above threshold.
\end{abstract}


The matter of photon number fluctuations in a single mode laser has been
studied more than thirty years ago in a number of pioneering works 
\cite{Hak64,Scu67,Lax69,Ris70,Hak72}.
At that time the question of photon number fluctuations at the laser 
threshold was resolved by adiabatically eliminating the excitation dynamics 
of the gain medium. That procedure allows the formulation of a photon 
number rate
equation for the single mode light field which can easily be solved 
analytically \cite{Wal94}.
However, the requirement that at every instant the gain function immediately
adjusts to the photon number in the cavity is not necessarily a valid 
assumption close to the laser threshold. 
In the following we show that the validity of the assumption depends 
on laser size and breaks down as the lasers get smaller. 
In particular, We derive a general expression for the photon number
fluctuations of both small and large lasers and demonstrate
that, indeed, near threshold the photon statistics of small lasers such as
typical semiconductor laser diodes are quite different from those
of larger lasers.

The dynamics of a single mode laser can be described by the rate equations
\begin{eqnarray}
\label{eq:energyflow1}
\frac{d}{dt} N &=& j-\frac{1}{\tau_{sp}} N - 2\frac{\beta}{\tau_{sp}}
                   \left(N-N_T\right)n
\nonumber \\
\frac{d}{dt} n &=& 2\frac{\beta}{\tau_{sp}}\left(N-N_T\right) n 
                   - \frac{1}{\tau_{cav}}n + \frac{\beta}{\tau_{sp}}N.
\end{eqnarray}
The dynamical variables are the photon 
number $n$ in the cavity mode and the excitation number $N$ in the gain 
medium.
The physical properties of the laser device are
characterized by four device parameters. These are
the spontaneous relaxation rate  $\tau_{sp}^{-1}$ of the excitations, 
the spontaneous emission factor $\beta$ defined as the ratio between the 
spontaneous emission rate into the laser mode and the total  
spontaneous relaxation rate of the excitations, 
the photon lifetime $\tau_{cav}$
inside the optical cavity, and the excitation number $N_T$ 
in the gain medium at transparency.  
The pump rate is given by $j$. With respect to electrically 
pumped semiconductor laser diodes it will be referred to as the injection
current.

The excitation density at transparency $N_T/V$, the spontaneous emission
rate $\tau^{-1}_{sp}$ and the dependence of the spontaneous emission factor
$\beta$ on the volume $V$ of the cavity are properties of the gain medium.
The order of magnitude of the cavity lifetime $\tau_{cav}$ is also defined
by the gain medium since the cavity loss rate $(2\tau_{cav})^{-1}$ must
be lower than the maximal amplification rate $\beta N_T/\tau_{sp}$ to
achieve laser operation. Therefore the device properties in the rate
equations depend mainly on the material properties of the gain medium and on
cavity size. For typical semiconductor laser diodes the device parameters
are
\begin{mathletters}
\label{devpara}
\begin{eqnarray} 
\beta V         &\approx& 10^{-14}\mbox{cm}^{3} 
\\
\frac{N_T}{V}   &\approx& 10^{18} \mbox{cm}^{-3} 
\\
\tau_{spont}    &\approx& 3 \times 10^{-9} \mbox{s}
\\
\tau_{cav}         &>&   1.5 \times 10^{-13} \mbox{s}.  
\end{eqnarray}
\end{mathletters}
The very fact that semiconductor lasers can be as small as a few $\mu$m in size is
a direct consequence of the relatively high spontaneous emission rate $\tau^{-1}_{sp}$. 

The stable stationary excitation number $\bar{N}$ and the stable
stationary photon number $\bar{n}$ may be obtained as a function of injection
current $j$. Transparency is reached at an injection current of
$j=N_T/\tau_{sp}$. The photon number at transparency is
\begin{equation}
n_T = \beta N_T \frac{\tau_{cav}}{\tau_{sp}}.
\end{equation}
This photon number is a measure of the cavity lifetime in units of
$\beta N_T/\tau_{sp}= 3\times 10^{-13}$ s. typical values will be between
one and two photons corresponding to cavity lifetimes of two to four times
the minimum required to achieve laser operation.

The laser threshold is defined by the light-current characteristic 
\begin{equation}
\label{eq:inject}
\frac{\bar{n}}{\tau_{cav}} = \frac{j-j_{th}-\tau_{cav}^{-1}}{2}
  +\frac{1}{2}\sqrt{(j-j_{th})^2+ 
                  \tau_{cav}^{-1}\left(2 j_{th}+\tau_{cav}^{-1}\right)}.
\end{equation}
via the threshold current $j_{th}$.
The threshold current $j_{th}$ marks the point at which
the transition from an almost negingible slope of the light-current 
characteristic (\ref{eq:inject}) to a slope of one takes place. 
This clearly corresponds to the intuitive notion of the laser light
``turning on'' at the laser threshold. 
In terms of the device parameters (\ref{devpara}) the threshold current $j_{th}$ reads
\begin{equation}
j_{th}= \frac{N_T}{\tau_{sp}}
\left(\left(1+\frac{1}{2 n_T}\right)
      -\beta \left(1+\frac{1}{n_T}\right)\right),
\end{equation}
where the cavity lifetime $\tau_{cav}$ has been expressed in terms of the 
photon number at transparency $n_T$. Note that since lasing requires
that $n_T>0.5$ this current is always less than twice the current 
required to reach transparency. It is thus possible to estimate the 
spontaneous emission factor directly from the threshold current.
In electrically pumped semiconductor laser diodes the product of the 
threshold current and the spontaneous emission factor
is approximately $0.5 \mu$A, e.g. a typical spontaneous emission factor 
of $10^{-5}$ corresponds to a threshold current of $50$ mA.
 
Since for the purpose of photon statistics we will in the following 
express the point of operation in terms of the average photon number 
$\bar{n}$ in the cavity, it is the photon number $n_{th}$ at 
$j=j_{th}$ which defines the laser threshold.
Assuming that 
the spontaneous emission factor $\beta$ is sufficiently smaller than one
this photon number reads
\begin{equation}
\bar{n}_{th} = 
\sqrt{\frac{n_T+\frac{1}{2}}{2\beta}}.
\end{equation}
This photon number is much higher than the photon number at 
transparency $n_T$, indicating that even below threshold
stimulated processes contribute more to the light field intensity
than spontaneous emissions.

With these definitions, the photon number fluctuations may now
be obtained from
the lineraized dynamics of the excitation number fluctuation
$\delta N=N-\bar{N}$ and the photon number fluctuation $\delta n=n-\bar{n}$   
which read
\begin{equation}
\label{eq:fluctstab}
\frac{d}{dt}\left(\begin{array}{c}\delta N \\ \delta n \end{array}\right)
= - \left(\begin{array}{cc} 
\Gamma_N & r \omega_R \\
r^{-1} \omega_R & \gamma_n
\end{array}\right)
\left(\begin{array}{c}\delta N \\ \delta n \end{array}\right)
+{\bf q}(t),
\end{equation}
where $\gamma_n$ is the relaxation rate of the photon number fluctuation 
and $\Gamma_N$ is the relaxation rate of the excitation number fluctuation.
The coupling rate $\omega_R$ describes the rate at which the holeburning 
effect of a photon number fluctuation acts back on that fluctuation. 
The fluctuation ratio $r$ is a measure of the relative importance of 
photon number noise with respect to excitation number fluctuations.
In terms of the stationary photon number $\bar{n}$ and the
four device parameters $N_T, n_T, \beta$ and $\tau_{sp}$
the rates and the ratio read 
\begin{mathletters}
\begin{eqnarray}
\gamma_n &=& \tau_{sp}^{-1}\frac{\beta N_T}{n_T} 
             \frac{n_T+\frac{1}{2}}{\bar{n}+\frac{1}{2}}
\\
\Gamma_N &=& \tau_{sp}^{-1}\left(1+2\beta\bar{n}\right) 
\\
\omega_R &=& \tau_{sp}^{-1}\sqrt{2 \beta \frac{\beta N_T}{n_T}
\left(\bar{n}-n_T\right)}
\\
r &=& \sqrt{\frac{N_T}{2 n_T} \frac{\left(\bar{n}-n_T\right)}
       {\left(\bar{n}+\frac{1}{2}\right)^2}}
.
\end{eqnarray}
\end{mathletters}
The fluctuation term $\bf{q}(t)$
is the shot noise arising from the quantization of excitation energy and
light field intensity. Since the excitation number $N_T$ at transparency is 
usually much larger than the average photon number $\bar{n}$ in the cavity, 
the ratio $r$ is much larger than one,
indicating that the fluctuations in the excitation number are much smaller
than the fluctuations in the photon number. It is therefore
reasonable to consider only the photon number contribution. Thus, 
\[
{\bf q}(t) = \left(\begin{array}{c} 0 \\ q_n \end{array} \right),
\]
\begin{equation}
\mbox{with} \quad 
\langle q_n(t)q_n(t+\Delta t)\rangle =
2 \bar{n}\left(\bar{n}+1\right) \gamma_n \; \delta (\Delta t).
\end{equation}
Note that this approximation is not valid for the low frequency part
of the noise spectrum since energy conservation requires that the 
low frequency noise is a function of the noise in the injection current
at high quantum efficiencies \cite{Yam86}.

With these assumptions we obtain the photon number fluctuations
\begin{equation}
\langle \delta n^2 \rangle =
\bar{n}\left(\bar{n}+1\right)\frac{1}{1+\frac{\Gamma_N \omega_R^2}
{\gamma_n \left(\omega_R^2 +\Gamma_N\gamma_n + \Gamma_N^2\right)}}. 
\end{equation}
This function is always lower than the thermal noise limit of 
$\langle \delta n^2 \rangle = \bar{n}(\bar{n}+1)$.
If the noise threshold is defined as the point at which
the photon number fluctuation drops below half the thermal noise 
limit then this threshold is determined from
\begin{equation}
\label{eq:condition}
\Gamma_N \omega_R^2 = \gamma_n 
        \left(\omega_R^2+\Gamma_N\gamma_n+\Gamma_N^2\right).
\end{equation}
If $\Gamma_N \gg \gamma_n$ and $\Gamma_N \gg \omega_R$, the excitation 
dynamics may be adiabatically eliminated. This is the basic assumption 
made in the conventional derivation of threshold noise \cite{Wal94}. 
Indeed one finds that the photon number
$\bar{n}_{1/2}$ at which the fluctuations correspond to one half thermal noise 
is in that case given by
\begin{equation}
\bar{n}_{1/2}=\sqrt{\frac{n_T+\frac{1}{2}}{2\beta}} = \bar{n}_{th}. 
\end{equation}
The noise threshold is then identical to the laser threshold 
in agreement with the predictions and observations made in the early days
of laser physics \cite{Hak64,Scu67,Lax69,Ris70,Hak72}.
However, the requirement that $\Gamma_N \gg \omega_R$ at 
$\bar{n}=\bar{n}_{th}$ is only valid for
\begin{equation}
\beta \ll \frac{1}{2n_T+1}\left(\frac{n_T}{\beta N_T}\right)^2 \approx 10^{-8}.
\end{equation}
In electrically pumped semiconductor lasers this would correspond to a threshold current
of more than 50~A. Therefore the assumption that the 
fluctuations $\delta N$ in excitation number can be adiabatically
eliminated at threshold does not apply to typical semiconductor laser diodes
which commonly have threshold currents significantly below 50 A.
Thus, for such devices the conventional theory no longer describes the photon 
number fluctuations at threshold. Instead, the complete dynamics 
of the fluctuations both in excitation number and in the number of photons
needs to be taken into account.

For typical semiconductor lasers with spontaneous emission factors
$\beta \gg 10^{-8}$ the coupling rate $\omega_R$ is 
much larger than the relaxation rate $\Gamma_N$ at laser threshold.
The noise threshold condition (\ref{eq:condition})
then reduces to
\begin{equation}
\Gamma_N=\gamma_n.
\end{equation}
As long as the total rate of spontaneous emission $\tau_{sp}^{-1}$ is
still greater than the rate of stimulated emission 
$2\beta \bar{n}\tau_{sp}^{-1}$, the photon number at the noise threshold is 
fixed by the properties of the gain medium at
\begin{equation}
\bar{n}_{1/2} = \beta N_T \left(1+\frac{1}{2 n_T}\right) \approx 10^4.
\end{equation}
However, stimulated emission takes over as the major relaxation
mechanism in the gain medium at an injection current
of twice the threshold current. 
The noise threshold is located beyond an injection current of two
times threshold current in laser devices with 
\begin{equation}
\beta > \left(2 \beta N_T \left(1+\frac{1}{2 n_T}\right)\right) 
      \approx 10^{-4}.
\end{equation}
This situation should apply in diodes with threshold currents of
less than 5 mA. In such devices the noise threshold is given by
\begin{equation}
\bar{n}_{1/2}= \sqrt{\frac{N_T}{2}\left(1+\frac{1}{2 n_T}\right)} 
\approx 10^2 \sqrt{\beta}.
\end{equation}

\setlength{\unitlength}{0.7 pt}
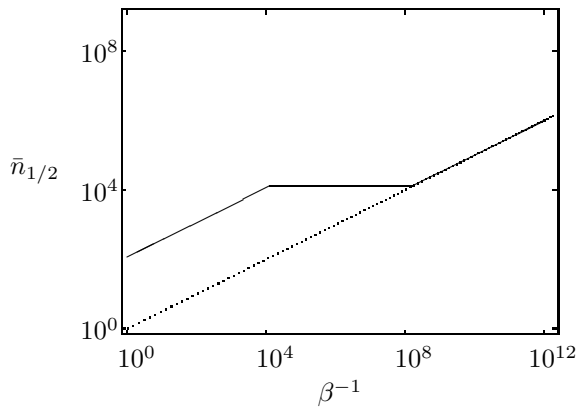
\begin{figure}
\begin{picture}(325,225)
\put(67,47){\line(1,0){236}}
\put(67,47){\line(0,1){176}}
\put(67,223){\line(1,0){236}}
\put(303,47){\line(0,1){176}}
\put(67,50){\line(1,0){2}}
\put(300,50){\line(1,0){2}}
\put(50,45){\makebox(10,10){$10^0$}}
\put(67,125){\line(1,0){2}}
\put(300,125){\line(1,0){2}}
\put(50,120){\makebox(10,10){$10^4$}}
\put(67,200){\line(1,0){2}}
\put(300,200){\line(1,0){2}}
\put(50,195){\makebox(10,10){$10^8$}}
\put(15,130){\makebox(10,10){$\bar{n}_{1/2}$}}
\put(70,47){\line(0,1){2}}
\put(70,220){\line(0,1){2}}
\put(70,30){\makebox(10,10){$10^0$}}
\put(145,47){\line(0,1){2}}
\put(145,220){\line(0,1){2}}
\put(145,30){\makebox(10,10){$10^4$}}
\put(220,47){\line(0,1){2}}
\put(220,220){\line(0,1){2}}
\put(220,30){\makebox(10,10){$10^8$}}
\put(295,47){\line(0,1){2}}
\put(295,220){\line(0,1){2}}
\put(295,30){\makebox(10,10){$10^{12}$}}
\put(180,10){\makebox(10,10){$\beta^{-1}$}}
\put(147,127){\line(-2,-1){77}}
\put(224,127){\line(-1,0){77}}
\put(224,127){\line(2,1){73}}
\bezier{100}(70,50)(220,125)(300,165)
\end{picture}
\caption{Noise threshold $\bar{n}_{1/2}$ as a function of the
spontaneous emission factor $\beta$.
The dotted line shows the threshold photon number $n_{th}$.}
\label{nhalf}
\end{figure}

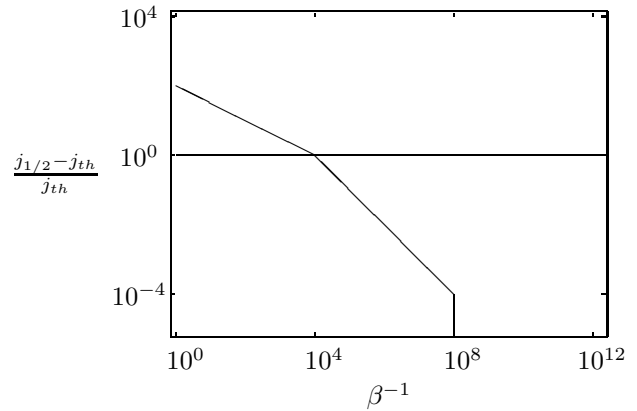
\begin{figure}
\begin{picture}(325,225)
\put(67,47){\line(1,0){236}}
\put(67,47){\line(0,1){176}}
\put(67,223){\line(1,0){236}}
\put(303,47){\line(0,1){176}}
\put(67,70){\line(1,0){2}}
\put(300,70){\line(1,0){2}}
\put(45,65){\makebox(10,10){$10^{-4}$}}
\put(67,145){\line(1,0){2}}
\put(300,145){\line(1,0){2}}
\put(45,140){\makebox(10,10){$10^0$}}
\put(67,220){\line(1,0){2}}
\put(300,220){\line(1,0){2}}
\put(45,215){\makebox(10,10){$10^4$}}
\put(0,130){\makebox(10,10){$\frac{j_{1/2}-j_{th}}{j_{th}}$}}
\put(70,47){\line(0,1){2}}
\put(70,220){\line(0,1){2}}
\put(70,30){\makebox(10,10){$10^0$}}
\put(145,47){\line(0,1){2}}
\put(145,220){\line(0,1){2}}
\put(145,30){\makebox(10,10){$10^4$}}
\put(220,47){\line(0,1){2}}
\put(220,220){\line(0,1){2}}
\put(220,30){\makebox(10,10){$10^8$}}
\put(295,47){\line(0,1){2}}
\put(295,220){\line(0,1){2}}
\put(295,30){\makebox(10,10){$10^{12}$}}
\put(180,10){\makebox(10,10){$\beta^{-1}$}}
\put(70,145){\line(1,0){230}}
\put(220,70){\line(0,-1){20}}
\put(220,70){\line(-1,1){75}}
\put(145,145){\line(-2,1){75}}
\end{picture}
\caption{Normalized difference between the injection current at one half
thermal noise $j_{1/2}$ and the threshold current $j_{th}$ as a function
of the spontaneous emission factor $\beta$.}
\label{jhalf}
\end{figure}

For semiconductor laser diodes the dependence of the photon number
at the noise threshold $\bar{n}_{1/2}$ on the spontaneous emission factor
$\beta$ may thus be summarized as illustrated in Fig.~\ref{nhalf}:
\begin{equation}
\bar{n}_{1/2} = \left\{ \begin{array}{ccc}
\frac{1}{\sqrt{\beta}} & \mbox{for} & \beta<10^{-8} \\
10^4 & \mbox{for} & 10^{-8}<\beta<10^{-4} \\
\frac{10^2}{\sqrt{\beta}} & \mbox{for} & 10^{-4}<\beta 
\end{array}\right. .
\end{equation}
Since the photon number at $j=2 j_{th}$ 
is approximately equal to $(n_T+1/2)/\beta$, the current
$j_{1/2}$ at which the photon number fluctuations drop to one-half 
their thermal value is approximately given by
\begin{equation}
\frac{j_{1/2}-j_{th}}{j_{th}}= \left\{ \begin{array}{ccc}
0 & \mbox{for} & \beta<10^{-8} \\
10^4 \beta & \mbox{for} & 10^{-8}<\beta<10^{-4} \\
10^2 \sqrt{\beta} & \mbox{for} & 10^{-4}<\beta 
\end{array}\right. 
\end{equation}
as illustrated in Fig.~\ref{jhalf}.

In conclusion, 
the assumption that above threshold the photon number 
fluctuations of laser light are lower than the fluctuations in 
equally coherent thermal light sources
is not valid in typical semiconductor lasers. In particular,
laser diodes with a threshold current of less than 5 mA still
fluctuate thermally far above threshold. Thus it is not possible
to distinguish in principle between lasers and thermal light sources 
based on the statistical properties of the emitted light field.
Therefore ``black box'' laser definitions disregarding the nature of the
internal light-matter interaction by which the light field is generated
do not apply to typical semiconductor laser diodes. 
If the definition of laser light is nevertheless based on the photon 
number fluctuations as suggested
e.g.~by Wiseman \cite{Wis97}, then the light from most laser
diodes could not be considered laser light even though it is definitely
generated by laser amplification. 
Moreover, a laser definition based on 
the condition that the relaxation rate of the excitations given 
by $\Gamma_N$ must be larger than the optical relaxation rate $\gamma_n$
entirely fails to relate to the original meaning of the 
acronym laser, i.e.~{\em light amplification by stimulated emission of radiation}.

It therefore seems to be reasonable to distinguish between a thermal
laser regime and a saturated laser regime separated by the noise threshold 
discussed above. In the thermal regime laser light indeed is indistinguishable 
from lamp light. In fact, the thermal laser regime naturally connects the 
saturated laser regime to the black body radiator from which the concepts
of spontaneous and stimulated emission originated \cite{Ein06}, thus 
providing a ``missing link'' in the theory of lasers and quantum optics.


%

\end{document}